\documentclass[12pt]{iopart}

\usepackage{graphicx}
\usepackage{subfigure}
\usepackage{caption}
\usepackage{upgreek}
\usepackage{dsfont}
\usepackage{color}
\usepackage{epstopdf}
\usepackage{amsfonts}
\usepackage{amssymb}
\usepackage{gensymb}
\begin{document}

\title{Controllable generation of mixed two-photon states}

\author{Jibo~Dai~$^1$~$^2$, Yink~Loong~Len~$^1$~$^2$~$^3$, Yong~Siah~Teo~$^4$, Leonid~A.~Krivitsky~$^1$\footnote{Author to whom any correspondence should be addressed.}, and Berthold-Georg~Englert~$^2$~$^3$}
\address{$^1$ Data Storage Institute, Agency for Science, Technology, and Research, 5 Engineering Drive I, Singapore 117608, Singapore}
\address{$^2$ Centre for Quantum Technologies, National University of Singapore, 3 Science Drive 2, Singapore 117543, Singapore}
\address{$^3$ Department of Physics, National University of Singapore, 2 Science Drive 3, Singapore 117542, Singapore}
\address{$^4$ Department of Optics, Palack{\'y} University, 17. listopadu 12, 77146 Olomouc, Czech Republic}
\ead{Leonid$\_$KRIVITSKIY@dsi.a-star.edu.sg}
\begin{abstract}
We report a controllable method for producing mixed two-photon states via Spontaneous Parametric Down-Conversion with a two-type-I crystal geometry. By using variable polarization rotators (VPRs), one obtains mixed states of various purities and degrees of entanglement depending on the parameters of the VPR. The generated states are characterized by quantum state tomography. The experimental results are found to be in good agreement with the theory. The method can be easily implemented for various experiments which require the generation of states with controllable degrees of entanglement or mixedness.
\end{abstract}
\maketitle


\section{Introduction}
Quantum information is a promising field that utilizes the nonclassical aspects of physical systems to perform sophisticated tasks of computations and communications. Quantum entanglement plays an important role in the implementation of these tasks. Arguably, the most famous entangled states are the Bell states, which are pure and maximally entangled. They are used in quantum cryptography \cite{PhysRevLett.67.661}, quantum teleportation \cite{bouwmeester1997experimental}, and the demonstration of various concepts of quantum mechanics \cite{Clauser, Aspect}. However, apart from the highly entangled pure states, there exists a vast uncharted region of state space, where states can be simultaneously mixed and entangled \cite{terhal2000bell,horodecki1996separability}. Mixed states are useful in investigations of quantum computing \cite{PhysRevLett.80.3408}, studies of the quantum-classical interface \cite{kwiat2004quantum}, and decoherence channels \cite{PhysRevA.72.052325}. These applications motivate an interest in the generation and characterization of mixed states. 

Up to now, several methods have been suggested for generating photonic mixed states. In Ref.$\,\,$\cite{ling2006preparation}, Werner states are produced with controlled addition of white noise to the Bell states. In Ref.$\,\,$\cite{PhysRevA.71.032329}, more sophisticated schemes for producing broad classes of states are presented. Such schemes employ an incoherent temporal mixing of state amplitudes, several decoherers, or a hybrid technique. However, as commented in Ref.$\,\,$\cite{PhysRevA.71.032329}, such schemes require many additional optical components, and are challenging in practice.

In this paper, we report a controllable method for producing mixed states, which requires only few additional optical components added to a Bell-state generation set-up. We use variable polarization rotators (VPRs), placed in the pump and signal beams of a conventional Spontaneous Parametric Down-Conversion (SPDC) set-up with two type-I crystals. Varying the parameters of the VPRs enables us to obtain mixed states of various purities and degrees of entanglement, which are then characterized by quantum state tomography. We remark that VPRs have been used earlier by Gogo \textit{et al.} in studies of quantum erasure \cite{PhysRevA.71.052103} without, however, presenting a detailed study and systematic characterization of the generated states.

\section{Mixed-state generation with VPR}
The basic principle underlying this method of producing mixed states is to generate incoherent mixtures of orthogonal Bell states with controllable weights. Consider a SPDC set-up with two type-I crystals with orthogonal axes, pumped by a laser polarized at $-45^\circ$ \cite{PhysRevA.60.R773}. The produced state is the Bell state $|\Phi_-\rangle = (|\mathrm{H_sH_i}\rangle - |\mathrm{V_sV_i}\rangle)/\sqrt{2}$, where H and V denote horizontal and vertical polarizations of down-converted photons, respectively, and the subscripts s-signal, and i-idler (hereafter omitted for simplicity) label the spatial modes. If the pump is polarized at $+45^\circ$, the orthogonal Bell state $|\Phi_+\rangle = (|\mathrm{HH}\rangle + |\mathrm{VV}\rangle)/\sqrt{2}$ is generated. By blending these two Bell states with different weights, one obtains the incoherent mixture
\begin{eqnarray}
\rho=\alpha|\Phi_-\rangle\langle\Phi_-| + (1-\alpha)|\Phi_+\rangle\langle\Phi_+| \\ \nonumber
=\frac{1}{2}\Big(|\mathrm{HH}\rangle\langle \mathrm{HH}|+|\mathrm{VV}\rangle\langle \mathrm{VV}|\Big)+\Big(\alpha-\frac{1}{2}\Big)\Big(|\mathrm{HH}\rangle\langle \mathrm{VV}|+|\mathrm{VV}\rangle\langle \mathrm{HH}|\Big),
\end{eqnarray}
where $\alpha$ ($0\leq\alpha\leq1$) is the weight of the state $|\Phi_-\rangle\langle\Phi_-|$ in the mixture.

To generate such mixed states, one switches the polarization of the pump back and forth between $-45^\circ$ and $+45^\circ$ and averages over a sufficient interval of time to sample both polarization states. Such a switching can be realized by inserting a VPR with electrically driven retardance in the pump beam. By applying a square waveform across the VPR, one achieves fast flippings between the two orthogonal linear polarizations. The parameter $\alpha$ is equal to the duty cycle ($\mathrm{DC}=\tau/T$, where $\tau$ is the duration of the high voltage, and $T$ is the period of the waveform) of the applied square waveform.

The purity $P=\mathrm{tr}\{\rho^2\}$ of such states is given by 
\begin{equation}
P=2\Big(\alpha-\frac{1}{2}\Big)^2+\frac{1}{2}.
\end{equation}

\noindent When $\alpha$ is zero or one, one of the Bell states is produced, and the purity is one. When $\alpha$ is 0.5, the two Bell states are mixed with the same proportion, and the purity has the minimum value of 0.5. Thus, changing $\alpha$ enables one to obtain states with different purities.

Different amounts of entanglement can also be obtained by varying the parameter $\alpha$. The tangle $T$ is a measure of quantum-coherence properties of a quantum state \cite{PhysRevLett.80.2245,PhysRevA.61.052306}. It has a value of zero for separable states, and one for maximally entangled states. For the states given by Eq.$\,\,$(1), the tangle is related to $\alpha$ by
\begin{equation}
T = (1-2\alpha)^2.
\end{equation}

\indent Generated states can be characterized by polarization correlation analysis. For the Bell states, the polarization correlation gives a visibility of 100$\%$ when measured in the $\pm45^\circ$ basis. For the states given by Eq.$\,\,$(1), the visibility varies with $\alpha$ according to
\begin{equation}
V = |1-2\alpha|=\sqrt{T}.\label{V}
\end{equation}
In particular, when $\alpha=0.5$, the generated state is $\rho=\frac{1}{2}\Big(|\mathrm{HH}\rangle\langle \mathrm{HH}|+|\mathrm{VV}\rangle\langle \mathrm{VV}|\Big)$, which results in zero visibility of the polarization correlation in $\pm45^\circ$ basis. Note that for the state given by Eq.$\,\,$(1), $V=\sqrt{T}$, however, this relation is not true in general.

With a single VPR in the pump beam and $\alpha$ set to 0.5, one produces mixed states with minimal purity $P=0.5$, see Eq.$\,\,$(2). However, for a completely mixed state, $P=1/d$, where $d$ is the dimension of the density matrix ($d=4$ for a two-photon state). The completely mixed state can be written as:
\begin{eqnarray}
\rho_{\mathrm{M}}=\frac{1}{4}\Big(|\mathrm{HH}\rangle\langle \mathrm{HH}|+|\mathrm{VV}\rangle\langle \mathrm{VV}|+|\mathrm{HV}\rangle\langle \mathrm{HV}|+|\mathrm{VH}\rangle\langle \mathrm{VH}|\Big)=\frac{\mathds{1}}{4}.
\end{eqnarray}
Thus, one has to include in the mixture the other two Bell states $|\Psi_-\rangle= (|\mathrm{HV}\rangle - |\mathrm{VH}\rangle)/\sqrt{2}$, and $|\Psi_+\rangle= (|\mathrm{HV}\rangle + |\mathrm{VH}\rangle)/\sqrt{2}$ with equal weights. These two Bell states can be generated in the same set-up by inserting a VPR with a half-wave retardance in either the signal or idler beam. In this case the VPR transforms the states $|\Phi_-\rangle\longrightarrow|\Psi_-\rangle$, and $|\Phi_+\rangle\longrightarrow|\Psi_+\rangle$. By switching between zero and half-wave retardance of the VPR, incoherent mixtures of $|\Phi_-\rangle\langle\Phi_-|$ and $|\Psi_-\rangle\langle\Psi_-|$, or $|\Phi_+\rangle\langle\Phi_+|$ and $|\Psi_-\rangle\langle\Psi_-|$, are produced.
Hence, by using two VPRs with duty cycles set to 0.5, one placed in the pump beam, and the other in the signal or idler beam, one can obtain the state given by Eq.$\,\,$(5). More generally, one can obtain mixtures of non-maximally entangled states $|\Psi_{\mathrm{ARB}}\rangle=\beta|$HH$\rangle+\gamma|$VV$\rangle$ by rotating the polarization of the pump beam to an arbitrary angle.

\section{Experimental Set-up}
\subsection{State Preparation}
\begin{figure}[b]
\includegraphics[width=155mm]{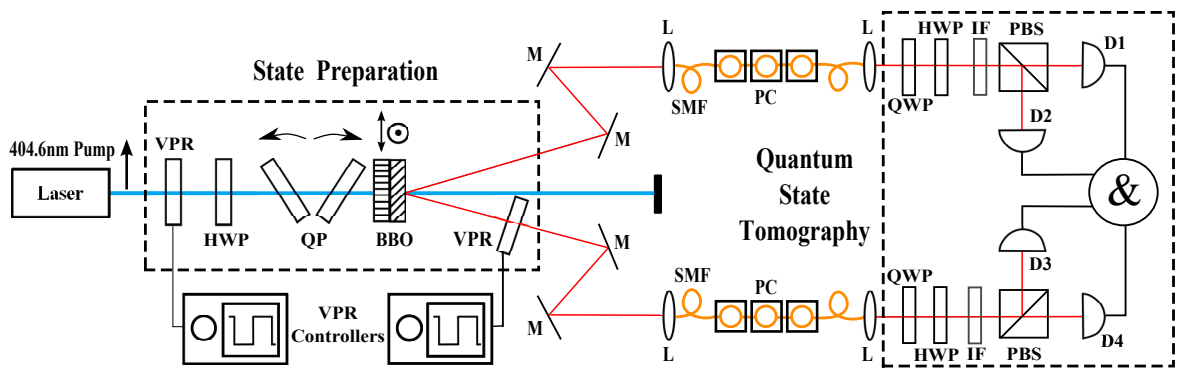}
\caption{Experimental set-up.  A cw diode laser pumps two type-I BBO crystals with orthogonal axes, where the SPDC occurs in the noncollinear frequency-degenerate regime. Mixed states are generated by inserting variable polarization rotators (VPRs) in the pump and signal beams. QP are quartz plates used to control the phase of the produced states. M are mirrors. The SPDC photons are coupled into single-mode fibers (SMF) with lenses (L). PC are polarization controllers, IF are interference filters. Quarter- and half-wave plates (QWP, HWP) and polarizing beam splitters (PBS) are used for quantum state characterization. D1-4 are single photon detectors, whose outputs are processed by a coincidence circuit (\&).}\label{fig1}
\end{figure}
The experimental set-up is shown in Fig.$\,\,$\ref{fig1}. Two beta-barium borate (BBO) crystals of $0.8\,\mathrm{mm}$ thickness with orthogonal optic axes are cut at $29.5^\circ$ for noncollinear frequency-degenerate phase matching. The BBOs are pumped by a frequency stabilized continuous wave diode laser at a wavelength of $404.6\,\mathrm{nm}$ (Ondax, LM series). The first crystal produces pairs of vertically polarized photons from the pump's horizontal polarization component, while the second crystal produces pairs of horizontally polarized photons from the pump's vertical polarization component. With noncollinear SPDC, the down-converted photons travel in different directions with a cone opening angle of $4^\circ$. Down-converted photons of narrow spatial bandwidth are collected into single-mode optical fibers (SMF) using aspherical lenses (L) with a focal length of $11\,\mathrm{mm}$, placed at a distance of $800\,\mathrm{mm}$ from the BBOs. The down-converted photons are filtered with interference filters (IF) centered at $810\,\mathrm{nm}$, with $10\,\mathrm{nm}$ full width at half maximum. Polarization rotations in the optical fibers are compensated for by manual polarization controllers (PC). 

With a half-wave plate (HWP) in the pump beam at $22.5^\circ$, the vertical polarization of the pump is rotated to $-45^\circ$, and the resulting state is $(|$HH$\rangle-e^{\mathrm{i}\phi}|$VV$\rangle)/\sqrt{2}$, where $\phi$ is the relative phase. This phase can be set to zero or integer multiples of $2\pi$ by manipulating the ellipticity of the pump with two quartz plates (QP) which are placed after the HWP, thus producing  $|\Phi_-\rangle = (|\mathrm{HH}\rangle - |\mathrm{VV}\rangle)/\sqrt{2}$.

Two different types of polarization rotators have been used in the experiment. The first one is a photoelastic modulator \textcolor{black}{(HINDS Instrument, I/FS50)}. It switches between zero and half-wave retardance at a frequency of $50\,\mathrm{kHz}$, with a fixed DC of 0.5. The second one is a liquid crystal retarder (LCR) \textcolor{black}{({Meadowlark, LRC-200)}}. It is configured so that zero retardance is obtained at the ``low" voltage of $V_\mathrm{L}=1.39\,\mathrm{V}$. The pump polarization entering the crystal remains at  $-45^\circ$, and $|\Phi_-\rangle$ is generated. For the ``high" voltage of $V_\mathrm{H}=1.64\,\mathrm{V}$, half-wave retardance is obtained. The pump polarization entering the crystal changes to $+45^\circ$, and $|\Phi_+\rangle$ is generated. A square waveform with the frequency of $1\,\mathrm{Hz}$ is applied to the LCR. The averaging is done over 3-minutes time intervals. By varying the duty cycle of the square waveform, different mixed two-photon states are obtained, see Eq.$\,\,$(1).

To produce the completely mixed state, given by Eq.$\,\,$(5), the photoelastic modulator is placed in the pump beam, and the LCR with $\mathrm{DC}=0.5$ is placed in the signal beam. The operation voltages of the LCR for down-converted photons are adjusted to $V_\mathrm{L}=1.84\,\mathrm{V}$ and $V_\mathrm{H}=6.95\,\mathrm{V}$. The frequency of the LCR square waveform is kept at $1\,\mathrm{Hz}$, and the averaging is done over 3-minutes time intervals.

\subsection{State characterization}
To fully characterize the states, quantum state tomography is performed. In both signal and idler arms, we install quarter-, and half-wave plates (QWP, HWP) and a polarizing beam splitter (PBS) \cite{tomo}. The transmission and reflection ports of both PBSs are directed to single-photon detectors \textcolor{black}{(Silicon Avalanche Photodiodes, quantum efficiency $\sim50\%$, Qutools Twin QuTD)}. Coincidence events are registered using the Time-To-Digital Converter \textcolor{black}{(quTAU)} with a time window of 5ns. The coincidences between any two of the detectors are recorded. By manipulating the wave plates in front of the PBSs, we perform measurements in nine different bases, which give an overcomplete measurement of 36 outcomes. Using the technique of maximum likelihood estimation \cite{paris2004quantum}, the state is inferred from the collected data.

\section{Results} 
\begin{figure}[t]
\begin{minipage}[b]{0.44\linewidth}
\includegraphics[width=1.05\textwidth]{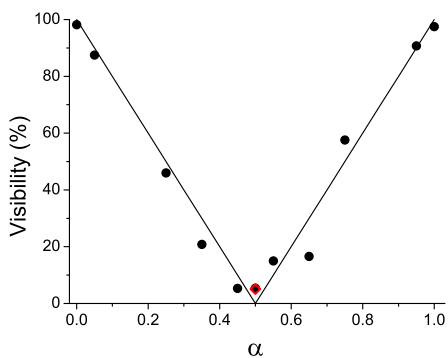}
\captionof{figure}{Dependence of the visibility of the polarization correlation measurements in the $\pm45^\circ$ basis on the DC of the LCR (solid circle), and of the photoelastic modulator (red open diamond). The solid line is the theoretical prediction. The error bars are smaller than the symbols.}
\label{vis}
\end{minipage}
\hspace{0.3cm}
\begin{minipage}[b]{0.56\linewidth}
\begin{tabular}[t]{ | c | c | p{5cm} |} 
    \hline
    $\alpha$ & Fidelity 
\\ \hline  
    0 & 0.9857$\pm$0.0005  \\ \hline
    0.05 & 0.9886$\pm$0.0005 \\ \hline
    0.25 & 0.9814$\pm$0.0006  \\ \hline
    0.35 & 0.9750$\pm$0.0012  \\ \hline
    0.45 & 0.9776$\pm$0.0007 \\ \hline
    0.50 & 0.9782$\pm$0.0006 \\ \hline
    \end{tabular}
\begin{tabular}[t]{ | c | c | p{5cm} |}
    \hline
    $\alpha$ & Fidelity
\\ \hline  
    0.55 & 0.9724$\pm$0.0012 \\ \hline
    0.65 & 0.9727$\pm$0.0008\\ \hline
    0.75 & 0.9792$\pm$0.0007 \\ \hline
    0.95 & 0.9786$\pm$0.0012 \\ \hline
    1 & 0.9785$\pm$0.0003\\ \hline \hline \hline
   \textcolor{blue}{PEM} & \textcolor{blue}{0.9890$\pm$0.0005}\\ \hline
   \textcolor{blue}{$\mathds{1}/4$} & \textcolor{blue}{0.9942$\pm$0.0003}\\ \hline
    \end{tabular}
    \captionof{table}{Fidelities of the reconstructed states with the target states for various DCs of the LCR. $F>97\%$ are consistently obtained for all the states. The last two entries with PEM and $\mathds{1}/4$ refer to the state generated using only the photoelastic modulator in the pump beam, and the completely mixed state generated using two VPRs, respectively. }
\label{tab}
\end{minipage}
\end{figure}

For the calibration of the set-up, the polarization correlation analysis in the $\pm 45^\circ$ basis is performed with the Bell state $|\Phi_-\rangle$. The constant ``low" voltage is applied across the LCR in the pump beam. The QWPs in the signal and idler arms are fixed at $0^\circ$. The HWP in the idler arm is fixed at $-22.5^\circ$ and the HWP in the signal arm is rotated. This causes the coincidence rate between the counts of two detectors in the transmitted ports of the PBSs to vary sinusoidally. The visibility is defined as $\displaystyle V=\Bigg|\frac{N_{\mathrm{+45^\circ}}-N_{\mathrm{-45^\circ}}}{N_{\mathrm{+45^\circ}}+N_{\mathrm{-45^\circ}}}\Bigg|$, where $N_\mathrm{+45^\circ}$ and $N_{\mathrm{-45^\circ}}$ are the coincidence rates when the HWP in the signal arm is oriented at $+22.5^\circ$ and $-22.5^\circ$, respectively.  The obtained value of visibility $97.3\pm1.4\%$ indicates high-quality of the produced Bell state. The analysis is then extended to the mixed states prepared with the LCR operating at different DCs, and with the photoelastic modulator. The obtained experimental visibilities are in good agreement with the theoretical expectation of Eq.$\,\,$(3), see Fig.$\,\,$\ref{vis}. One observes a degradation of the visibility as $\alpha$ increases from 0 to 0.5. After which, the visibility is gradually restored  as $\alpha$ increases from 0.5 to 1. 

Next, quantum state tomography on the generated states is performed. The matrix elements of reconstructed density matrices are shown graphically in Fig.$\,\,$3, for $\alpha$=0.05, 0.25, and 0.5. The fidelities, $F$, of the reconstructed density matrices with the states given by Eq.$\,\,$(1) are calculated. Using the definition $\displaystyle F=\mathrm{tr}\big\{|\sqrt{\rho}\sqrt{\sigma}|\big\}$, where $\rho$ is the maximum likelihood estimator, and $\sigma$ is the target state, fidelities $F>97\%$ are consistently observed, see Table.$\,\,$\ref{tab}. 

As an example, for $\alpha=0.25$, the reconstructed density matrix expressed in the HV basis is\\
$\rho_{_{ \alpha=0.25}}\widehat{=}\left(
\begin{array}{cccc}
 \hphantom{-}0.545\hphantom{-0.012i_i}  \,\,&  \hphantom{-} 0.049+0.012\mathrm{i}  \,\,&   -0.013+0.038\mathrm{i}  \,\,&   -0.232+0.110\mathrm{i} \\
 \hphantom{-} 0.049-0.012\mathrm{i}  \,\,&  \hphantom{-} 0.008\hphantom{+0.012i_i}  \,\,&   -0.005+0.008\mathrm{i}  \,\,&  \hphantom{-} 0.015+0.004\mathrm{i} \\
  -0.013-0.038\mathrm{i}  \,\,&   -0.005-0.008\mathrm{i}  \,\,&  \hphantom{-} 0.013\hphantom{+0.012i_i} \,\,&   -0.039+0.006\mathrm{i} \\
  -0.232-0.110\mathrm{i}  \,\,&  \hphantom{-} 0.015-0.004\mathrm{i}  \,\,&   -0.039-0.006\mathrm{i}  \,\,&  \hphantom{-} 0.434\hphantom{+0.012i_i}
\end{array}
\right).$\\
\noindent The reconstructed state has a tangle of $T=\,\,0.2476\pm 0.0049$, a purity of $P=0.6295\pm 0.0022$, and a fidelity with the target state of $F=0.9814\pm 0.0006$.

As one can see from Fig.$\,\,$3, an increase of $\alpha$ from 0 to 0.5 causes vanishing of the off-diagonal elements, which indicates the decreasing amount of entanglement of the mixed states. The calculated values of the tangle for the reconstructed states, are in good agreement with Eq.$\,\,$(3), see Fig.$\,\,$\ref{tangle}(a). The state with almost zero  tangle is generated with the photoelastic modulator. It performs better than the LCR at $\alpha$=0.5, as the wavelength of the pump laser is at the edge of the working range of the LCR.
\begin{figure}[t!]
\hfill
\subfigure{\includegraphics[width=5cm]{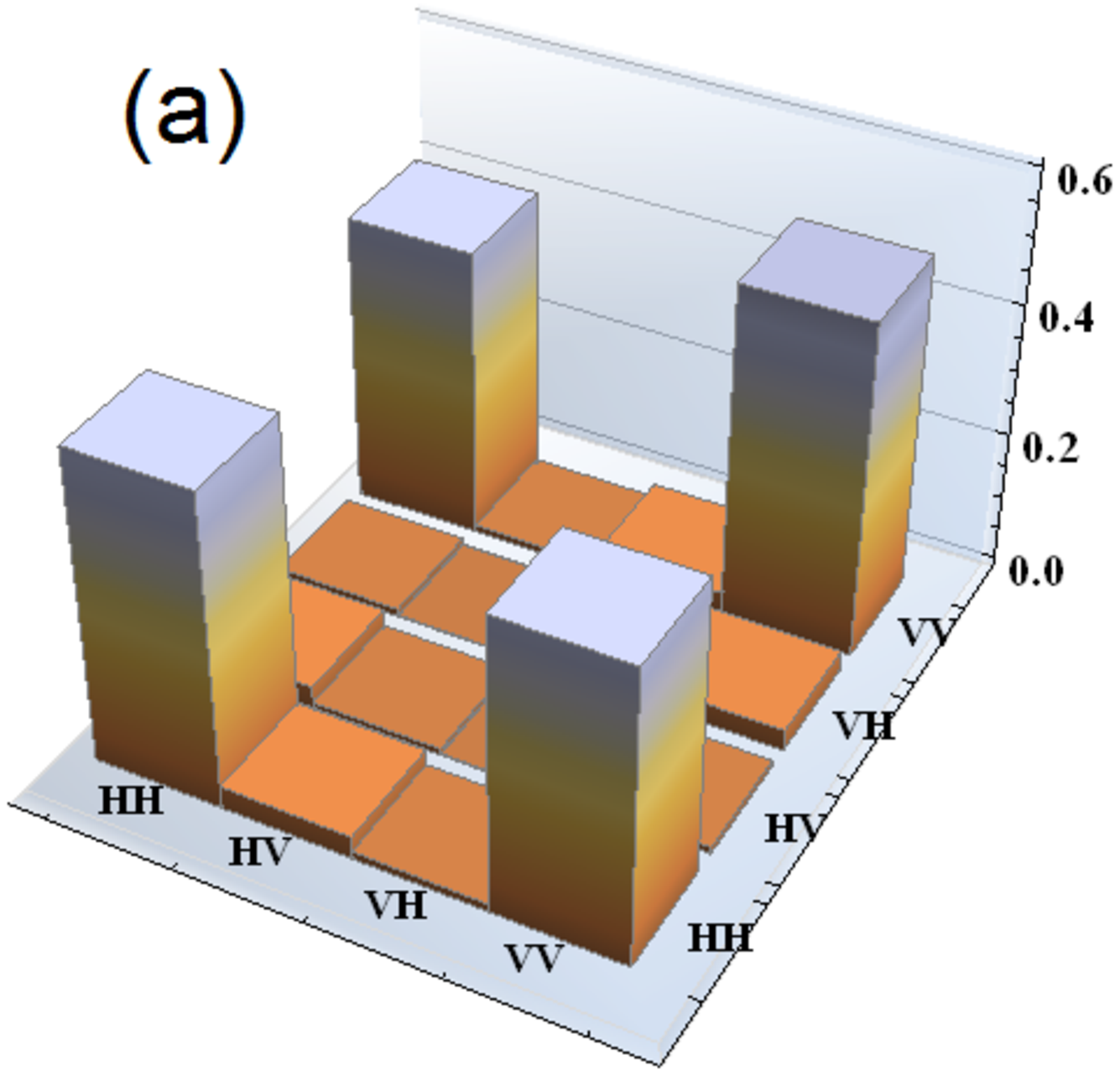}}
\hfill
\subfigure{\includegraphics[width=5cm]{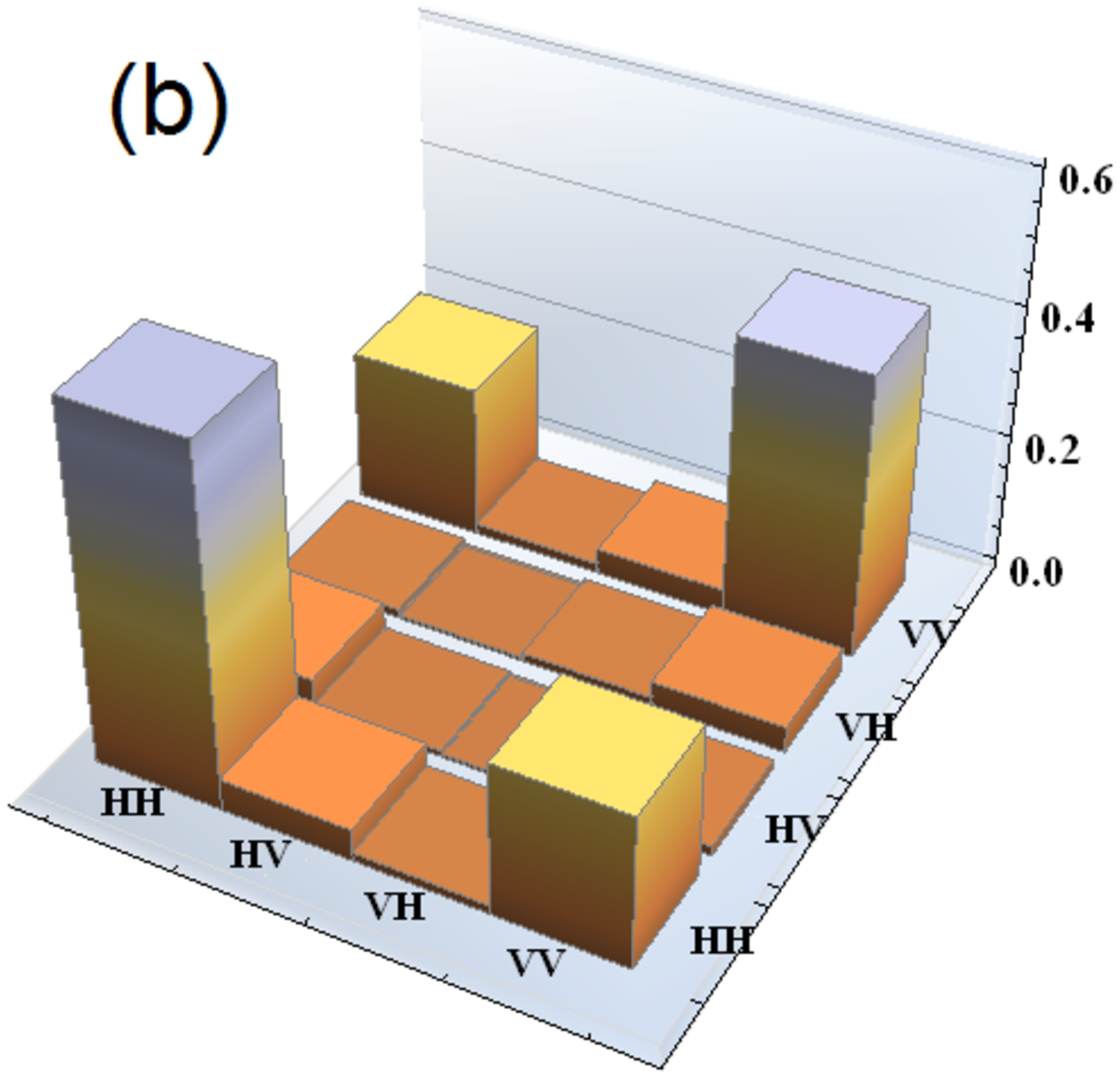}}
\hfill
\subfigure{\includegraphics[width=5cm]{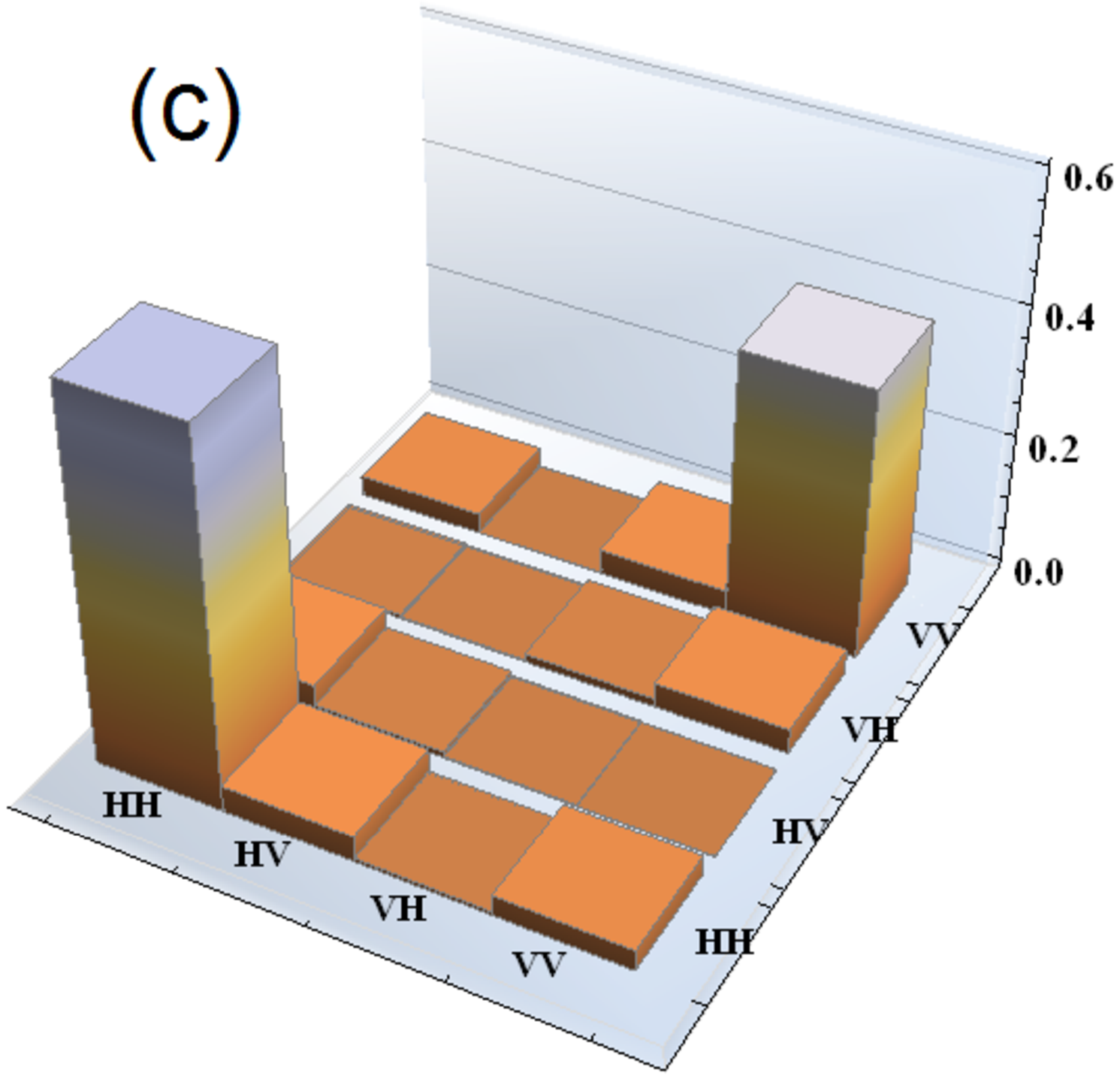}}
\hfill
\subfigure{\includegraphics[width=5cm]{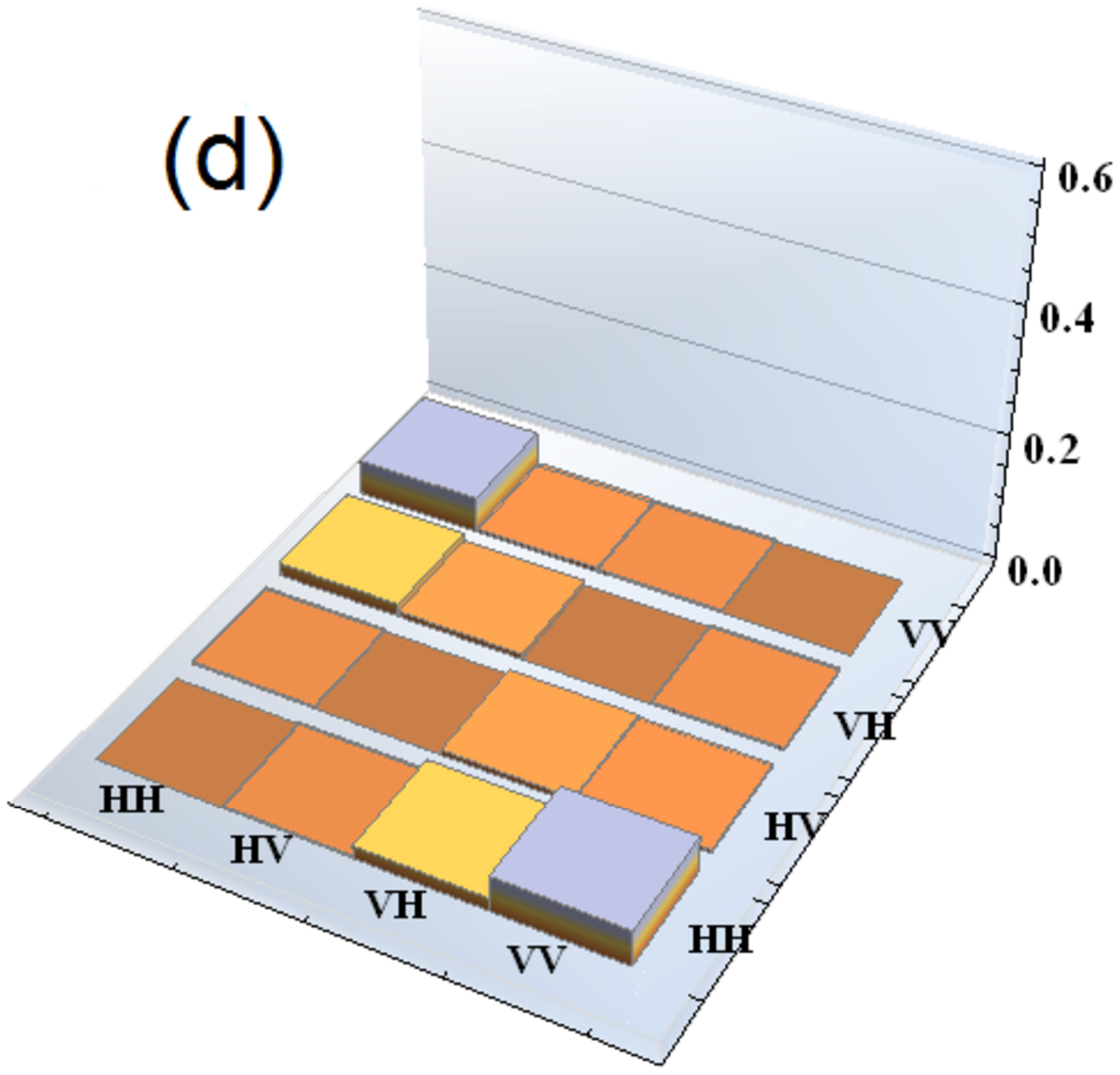}}
\hfill
\subfigure{\includegraphics[width=5cm]{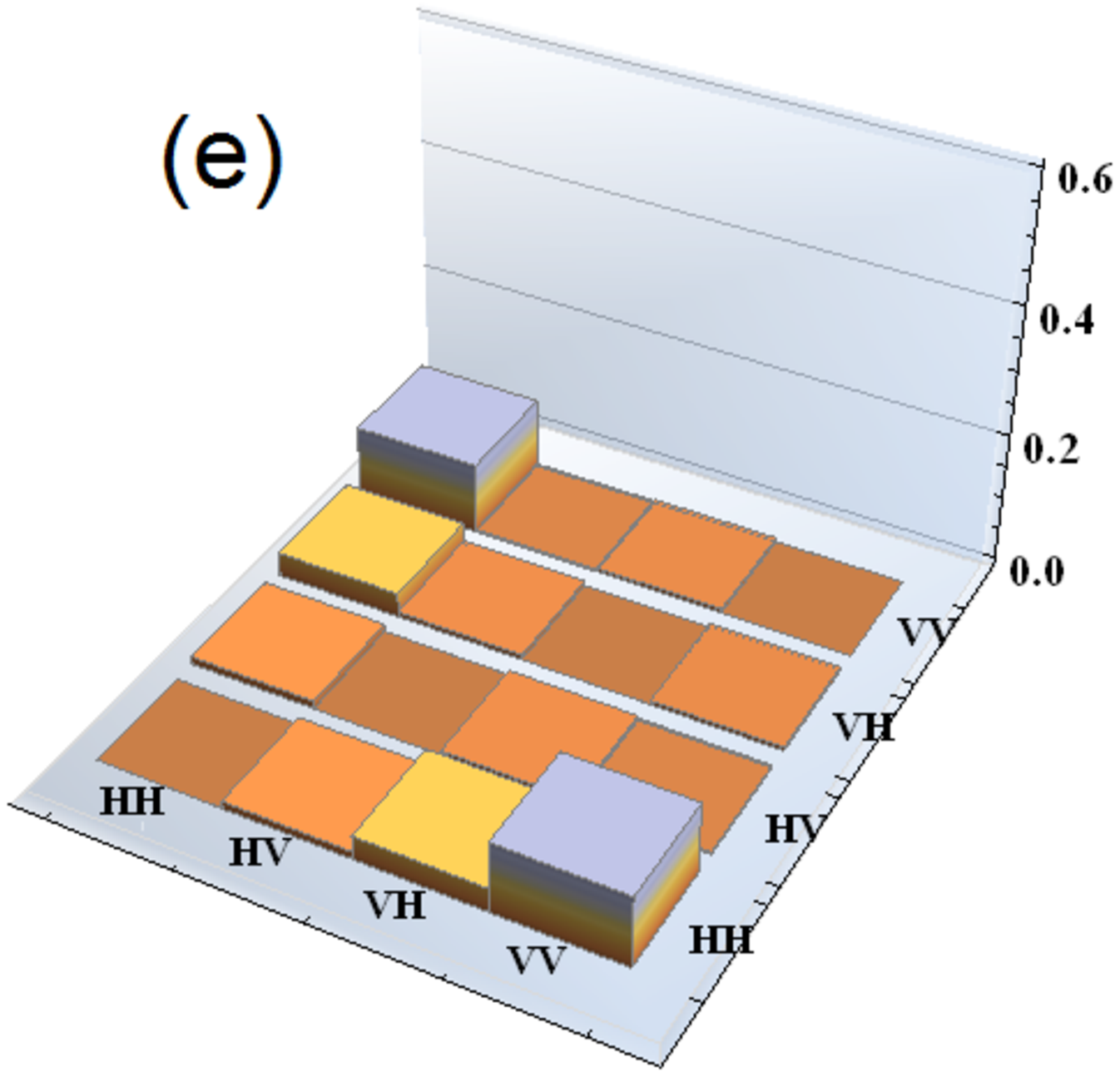}}
\hfill
\subfigure{\includegraphics[width=5cm]{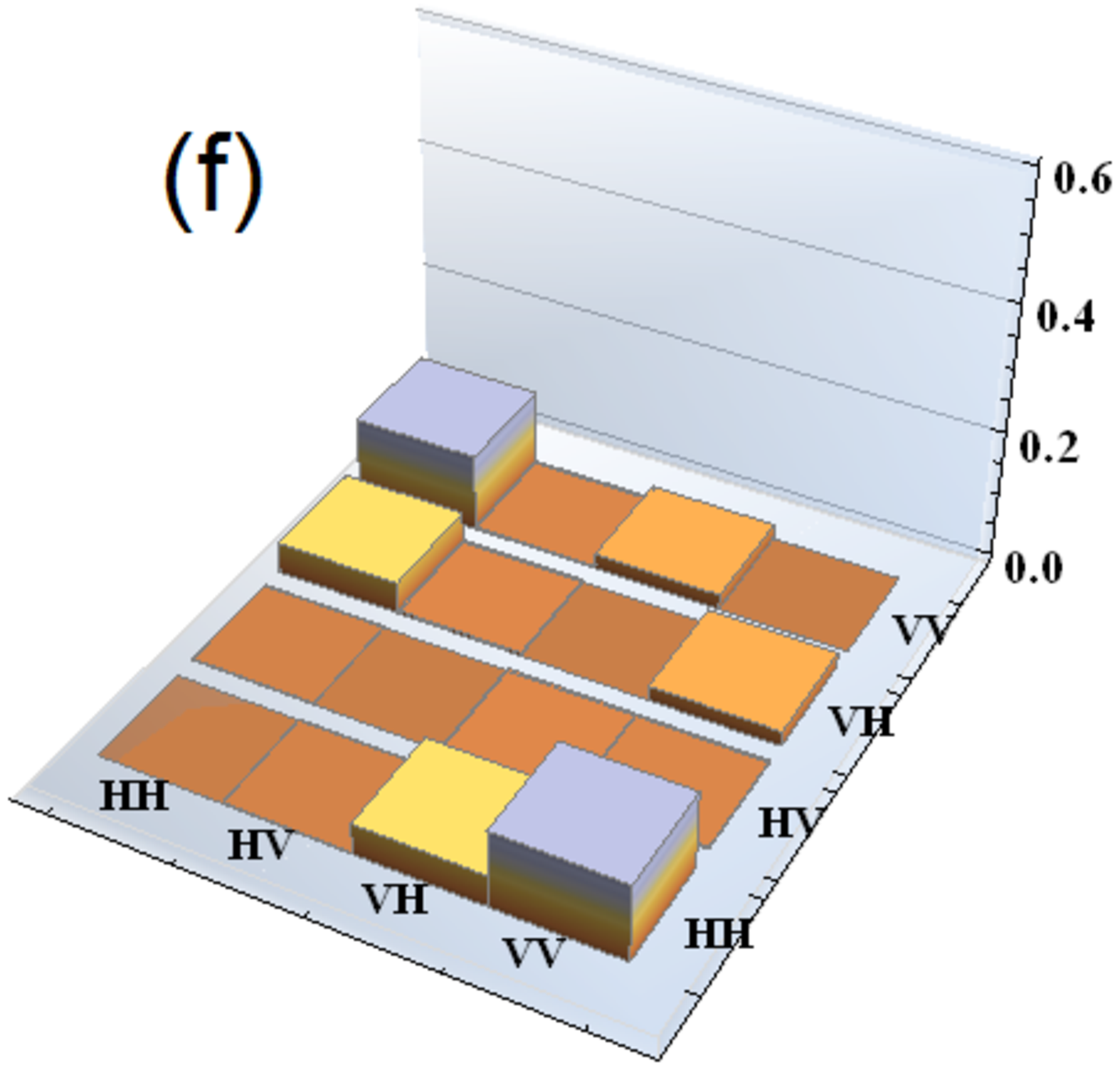}}
\hfill
\caption{Absolute values of  real (a, b, c) and  imaginary (d, e, f) parts of the density matrices representing the reconstructed states for  0.05DC (a, d);  0.25DC (b, e) and  0.50DC (c, f). The vanishing of the off-diagonal elements is clearly seen. }
\end{figure}
\begin{figure}[t!]
  	\centering    
  	\subfigure{\includegraphics[width=75mm]{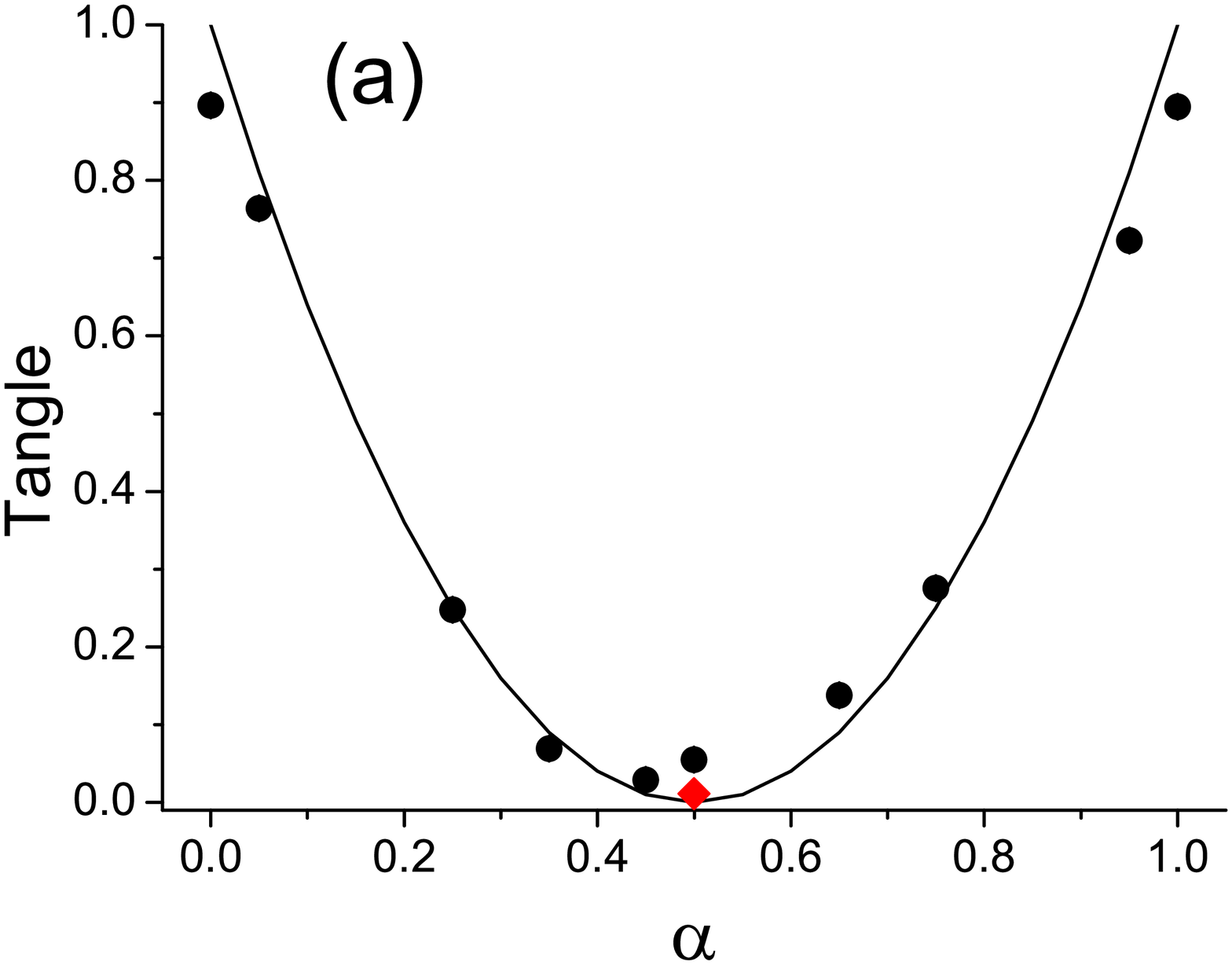}}
  	\subfigure{\includegraphics[width=75mm]{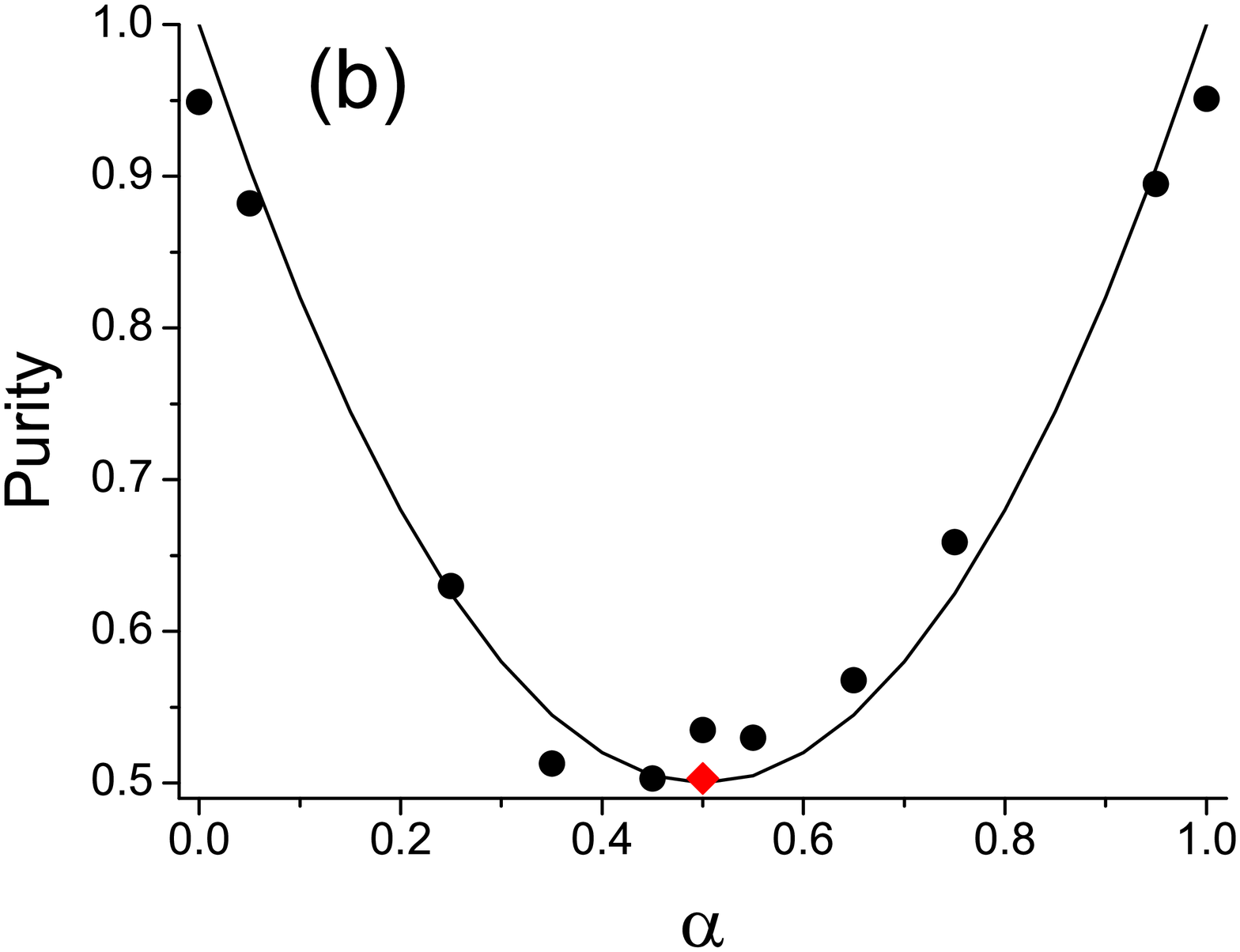}}
		\caption{Dependence of  tangle (a) and  purity (b) of reconstructed states on the duty cycle of the LCR (solid circles), and of the photoelastic modulator (red diamond). The solid curves are the theoretical predictions. The error bars are smaller than the symbols used for both figures.}\label{tangle}
\end{figure}

Lastly, one verifies that the method indeed produces states with different amounts of mixedness. The dependence of the purity of the states on $\alpha$ is shown in Fig.$\,\,$\ref{tangle}(b). The obtained results are close to those predicted by Eq.$\,\,$(2). 

\begin{figure}[t!] 
\centering
\subfigure{\includegraphics[width=5.5cm]{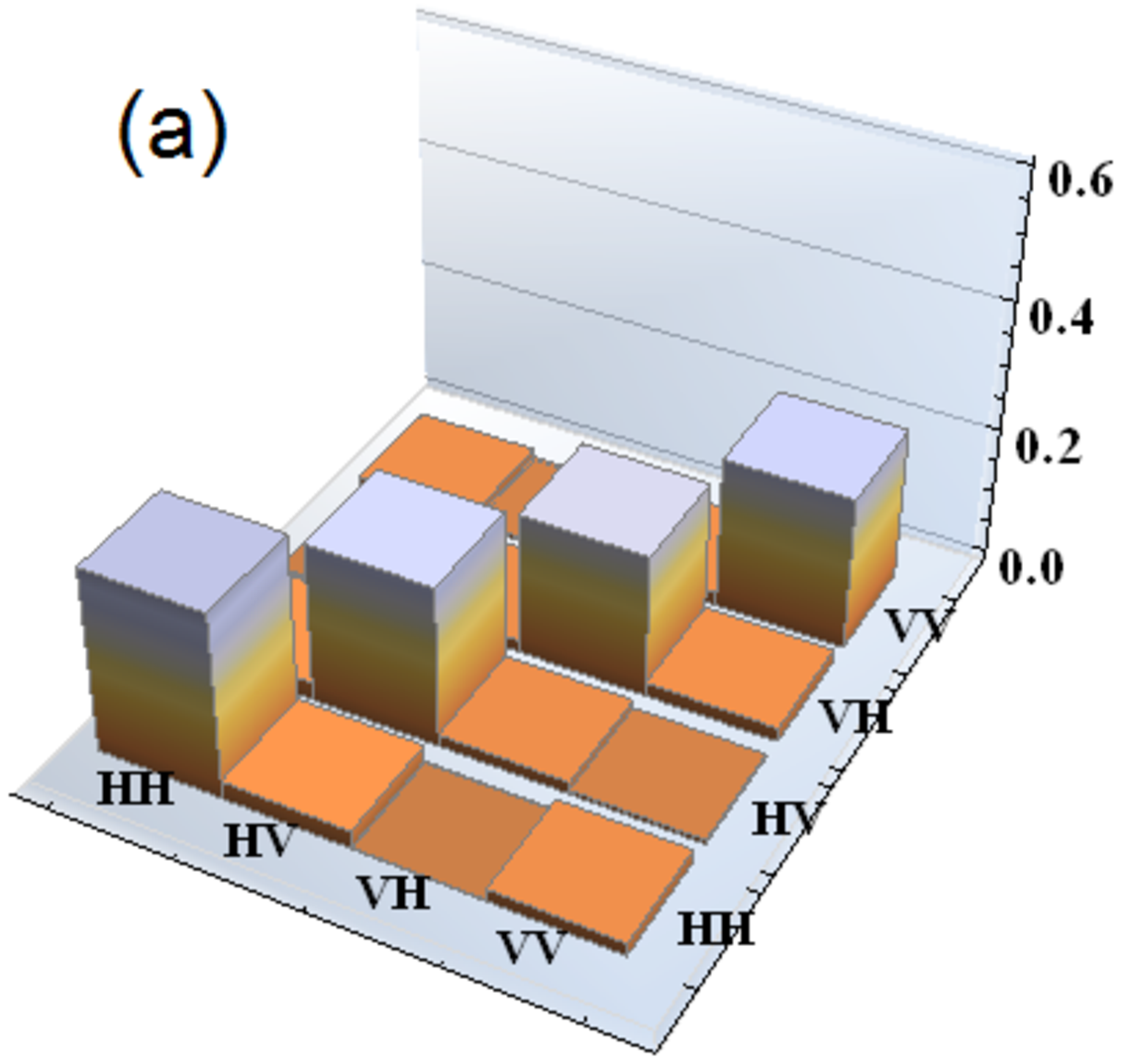}}
\hspace{0.4cm}
\subfigure{\includegraphics[width=5.5cm]{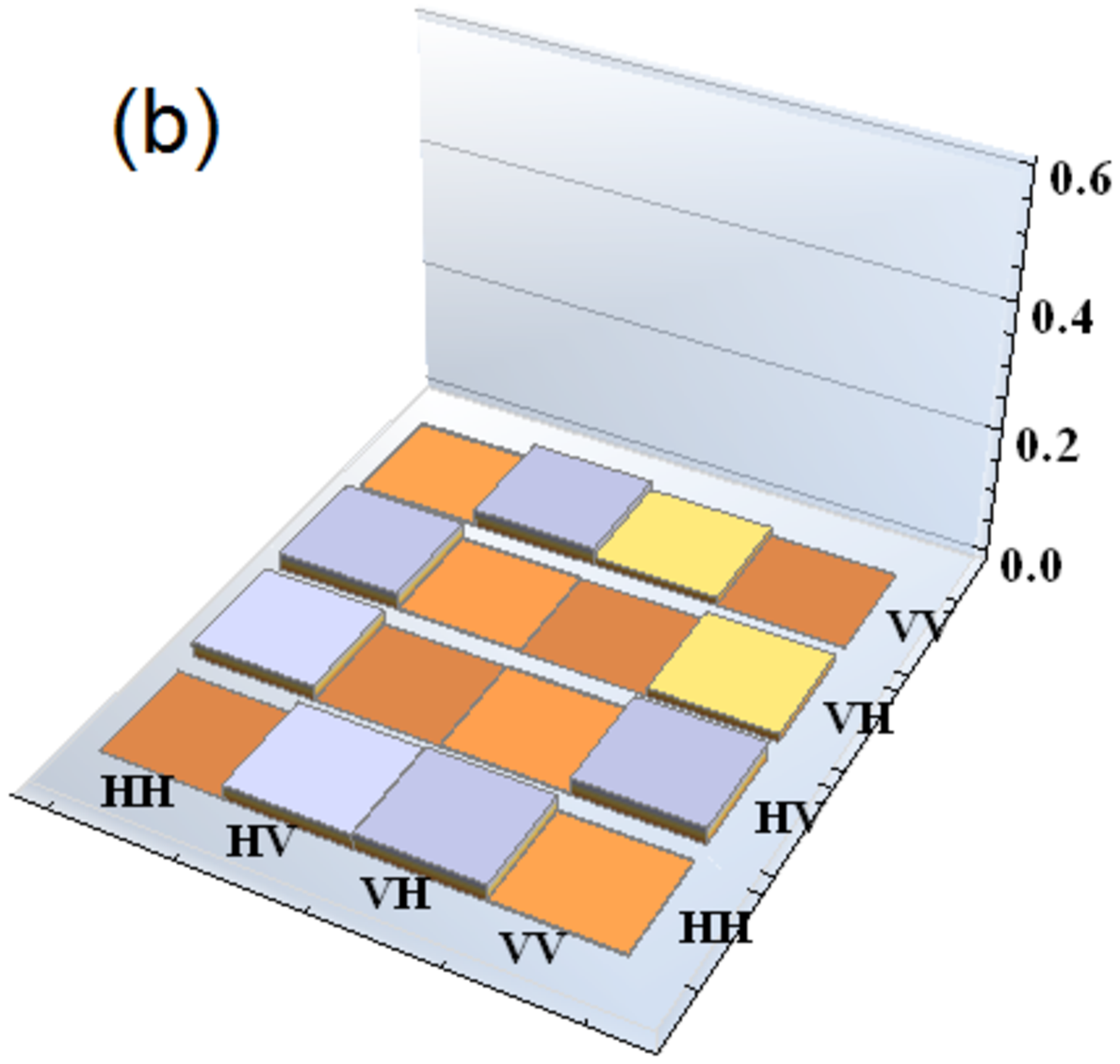}}
		\caption{Absolute values of  real (a) and  imaginary (b) parts of the reconstructed density matrix for the completely mixed state. }\label{cmix}
\end{figure}

To demonstrate generation of the completely mixed state,  the set-up is modified by inserting VPRs in the pump and in the signal beams, see Sec. 3.1. The reconstructed  density matrix of the state is shown graphically in  Fig.$\,\,$\ref{cmix}. The reconstructed state has a tangle of zero, a purity of $P=0.2615\pm 0.0007$, and a fidelity with the target state of $F=0.9942\pm 0.0003$. Such a state is separable and highly mixed. 
                              
\section{Conclusion}
In summary, an accessible way of generating controllable mixtures of the Bell states with high fidelities is reported. There is no need for extra optical components, except the VPRs. Furthermore, the method does not require any reconfiguration of the standard set-up for Bell states generation, as the VPRs can be easily turned on and off, to generate mixed and pure states, respectively.  In addition, it offers the important advantage of interferometric stability. The presented method will facilitate state generation and can be used in various quantum-information processing applications. We are planning to use it for an experimental realization of qubit-pair tomography with witness bases \cite{PhysRevA.81.052339}.

\ack
We thank Dmitry Kalashnikov for his assistance with the experiment and valuable discussions. This work is supported by the National Research Foundation and the Ministry of Education, Singapore, and is co-financed by the European Social Fund and the state budget of the Czech Republic, project No. CZ.1.07/2.3.00/30.0004 (POST-UP).


\section*{References}
\begin{thebibliography}{99}
\bibitem{PhysRevLett.67.661}
Ekert A K 1991 Quantum cryptography based on Bell's theorem {\it \PRL} {\bf 67} 661-3
\bibitem{bouwmeester1997experimental}
Bouwmeester D, Pan J-W, Mattle K, Eibl M, Weinfurter H and Zeilinger A 1997 Experimental quantum teleportation {\it Nature} {\bf390} 575-9
\bibitem{Clauser}
Clauser J F and Shimony A 1978 Bell's theorem. Experimental tests and implications {\it Rep. Prog. Phys.} {\bf 41} 1881
\bibitem{Aspect}
Aspect A, Grangier P and Roger G 1981 Experimental Tests of Realistic Local Theories via Bell's Theorem {\it \PRL} {\bf 47} 460-3
\bibitem{terhal2000bell}
Terhal B 2000 Bell inequalities and the separability criterion {\it \PL}A {\bf 271} 319-26
\bibitem{horodecki1996separability}
Horodecki M, Horodecki P and Horodecki R. 1996 Separability of mixed states: necessary and sufficient conditions {\it \PL}A {\bf 223} 1-8
\bibitem{PhysRevLett.80.3408}
Chuang I L, Gershenfeld N and Kubinec M 1998 Experimental implementation of fast quantum searching {\it \PRL} {\bf 80} 3408-11
\bibitem{kwiat2004quantum}
Kwiat P G and Englert B-G 2004 Quantum erasing the nature of reality or, perhaps, the reality of nature? {\it Science and Ultimate Reality: Quantum Theory, Cosmology, and Complexity} (Cambridge University Press) p 306-29
\bibitem{PhysRevA.72.052325} 
Ziman M and Bu{\v z}ek V 2005 Concurrence versus purity: Influence of local channels on Bell states of two qubits {\it\PR}A {\bf72} 052325
\bibitem{ling2006preparation}
Ling A, Han PY, Lamas-Linares A, and Kurtsiefer C 2006 Preparation of Bell states with controlled white noise {\it Laser Phys.} {\bf 16} 1140-4
\bibitem{PhysRevA.71.032329} 
Wei T-C, Altepeter J B, Branning D, Goldbart P M, James D F V, Jeffrey E, Kwiat P G, Mukhopadhyay S and Peters N A 2005 Synthesizing arbitrary two-photon polarization mixed states {\it\PR}A {\bf71} 032329
\bibitem{PhysRevA.71.052103} 
Gogo A, Snyder W D and Beck M 2005 Comparing quantum and classical correlations in a quantum eraser {\it\PR}A {\bf71} 052103
\bibitem{PhysRevA.60.R773} 
Kwiat P G, Waks E, White A G, Appelbaum I and Eberhard P H 1999 Ultrabright source of polarization-entangled photons {\it\PR}A {\bf60} R773-6
\bibitem{PhysRevLett.80.2245} 
Wootters W K 1998 Entanglement of formation of an arbitrary state of two qubits {\it\PRL} {\bf80} 2245-8
\bibitem{PhysRevA.61.052306} 
Coffman V, Kundu J and Wootters W K 2000 Distributed entanglement {\it\PR}A {\bf61} 052306
\bibitem{tomo}
James D F V, Kwiat P G, Munro W J and White A G 2001 Measurement of qubits {\it\PR}A {\bf64} 052312
\bibitem{paris2004quantum} 
Paris M and {\v R}eh{\'a}{\v c}ek J 2004 {\it Quantum state estimation}, Lecture Notes in Physics, vol. {\bf649} (Springer)
\bibitem{PhysRevA.81.052339} 
Zhu H, Teo Y S and Englert B-G 2010 Minimal tomography with entanglement witnesses {\it\PR}A {\bf81} 052339
\end{thebibliography}
\end{document}